\documentclass[aps,10pt,reprint,groupedaddress,superscriptaddress,nofootinbib]{revtex4-2}

\usepackage{graphicx}
\usepackage{upgreek}
\usepackage{braket}
\usepackage{siunitx}
\usepackage{enumitem}
\usepackage[normalem]{ulem}
\usepackage[hidelinks, colorlinks=true,allcolors=blue]{hyperref}
\usepackage{amsmath}
\newcommand{\hhbar}{\mathchar'26\mkern-9mu h}
\newcommand{\dL}{ d_\text{L} }
\newcommand{\thetaL}{ \theta_\text{L} }
\newcommand{\thetaD}{ \theta_\text{D} }
\newcommand{\Er}{ E_\text{r} }
\newcommand{\kL}{ k_\text{L} }

\begin{document}

\title{Tunable optical lattices for the creation of matter-wave lattice solitons}

\author{Robbie Cruickshank}
\affiliation{Department of Physics and SUPA, University of Strathclyde, Glasgow G4 0NG, United Kingdom }
\author{Arthur~La~Rooij}
\affiliation{Department of Physics and SUPA, University of Strathclyde, Glasgow G4 0NG, United Kingdom }
\author{Ethan F. Kerr}
\affiliation{Department of Physics and SUPA, University of Strathclyde, Glasgow G4 0NG, United Kingdom }
\author{Timon Hilker}
\affiliation{Department of Physics and SUPA, University of Strathclyde, Glasgow G4 0NG, United Kingdom }
\author{Stefan Kuhr}
\affiliation{Department of Physics and SUPA, University of Strathclyde, Glasgow G4 0NG, United Kingdom }
\author{Elmar Haller}
\affiliation{Department of Physics and SUPA, University of Strathclyde, Glasgow G4 0NG, United Kingdom }

\date{\today}

\begin{abstract}
We present experimental techniques that employ an optical accordion lattice with dynamically tunable spacing to create and study bright matter-wave solitons in optical lattices. The system allows precise control of lattice parameters over a wide range of lattice spacings and depths. We detail calibration methods for the lattice parameters that are adjusted to the varying lattice spacing, and we demonstrate site-resolved atom number preparation via microwave addressing. Lattice solitons are generated through rapid quenches of the atomic interaction strength and the external trapping potential. We systematically optimize the quench parameters, such as duration and final scattering length, to maximize soliton stability. Our results provide insight into nonlinear matter-wave dynamics in discretized systems and establish a versatile platform for the controlled study of lattice solitons.
\end{abstract}

\maketitle

\section{Introduction}

Solitons are localized, non-dispersive wave packets that preserve their shape during propagation due to a balance between dispersion and nonlinearity. First observed in shallow water waves by Russell in 1834 \cite{russell1845}, solitons have since been realized in a wide range of physical systems, including plasmas \cite{ikezi1973}, nonlinear optical media \cite{mollenauer1980,duree1993}, and biological cells~\cite{kuwayama2013}. In Bose–Einstein condensates (BECs), solitons arise as nonlinear matter-waves governed by the Gross–Pitaevskii equation with two prominent examples~\cite{Kartashov2011}.  
Firstly, bright matter-wave solitons, which are localized density peaks on a low-density background, have been subject of extensive experimental and theoretical investigations~\cite{abdullaev2005,parker2007}, studying their formation~\cite{bradley1997,khaykovich2002}, collapse dynamics~\cite{sackett1998,gerton2000,donley2001}, collisions~\cite{nguyen2014}, and interactions with external potentials, such as reflection and transmission at barriers~\cite{marchant2013,marchant2016}, among other phenomena~\cite{strecker2002,cornish2006}. Secondly,  dark matter-wave solitons are characterized by density dips accompanied by a phase slip in a higher-density background~\cite{jackson1998} and have been created~\cite{burger1999,denschlag2000} and observed interacting with other solitons~\cite{becker2008,weller2008}. Various techniques have been used to realize matter-wave solitons, including rapid or adiabatic changes in the atomic interaction strength~\cite{khaykovich2002,dicarli2019b,luo2020}, phase imprinting~\cite{burger1999,denschlag2000}, and the exploitation of modulational instabilities in the carrier medium~\cite{nguyen2017}.

Over the last decades, substantial theoretical research has been dedicated to investigating solitons in the presence of a periodic potentials~\cite{trombettoni2001, ahufinger2004, salasnich2007, maluckov2008}. These so-called lattice solitons have been identified in various physical systems, from molecular chains~\cite{davydov1973,kruglov1984} to nonlinear optical waveguides~\cite{eisenberg1998,morandotti1999,fleischer2003}, and quantum gases in optical lattices~\cite{trombettoni2001,ostrovskaya2003,eiermann2004}. They exist in both one and two dimensions~\cite{ahufinger2004,baizakov2003,morandotti1999,fleischer2003}, and exhibit intricate transport behavior~\cite{trombettoni2001, ahufinger2004, brazhnyi2011, franzosi2011}. We have recently reported the first observation of this novel type of soliton in quantum gases with attractive interactions \cite{cruickshank2025}. There, we were able to accurately prepare atoms in a defined number of lattice sites to demonstrate two types of lattice solitons: single-site and multi-site which span one and many sites respectively. To achieve this, a rapid quench from repulsive to attractive interactions was utilized, accompanied by the removal of axial confinement along the lattice direction. We found that the stability of the resulting lattice solitons depends critically on the overlap between the initial wavefunction and the target soliton density profile. Imperfect matching leads to breathing dynamics, while strong mismatch can result in collapse or dispersion.

\begin{figure}[t]
\includegraphics[width=\columnwidth]{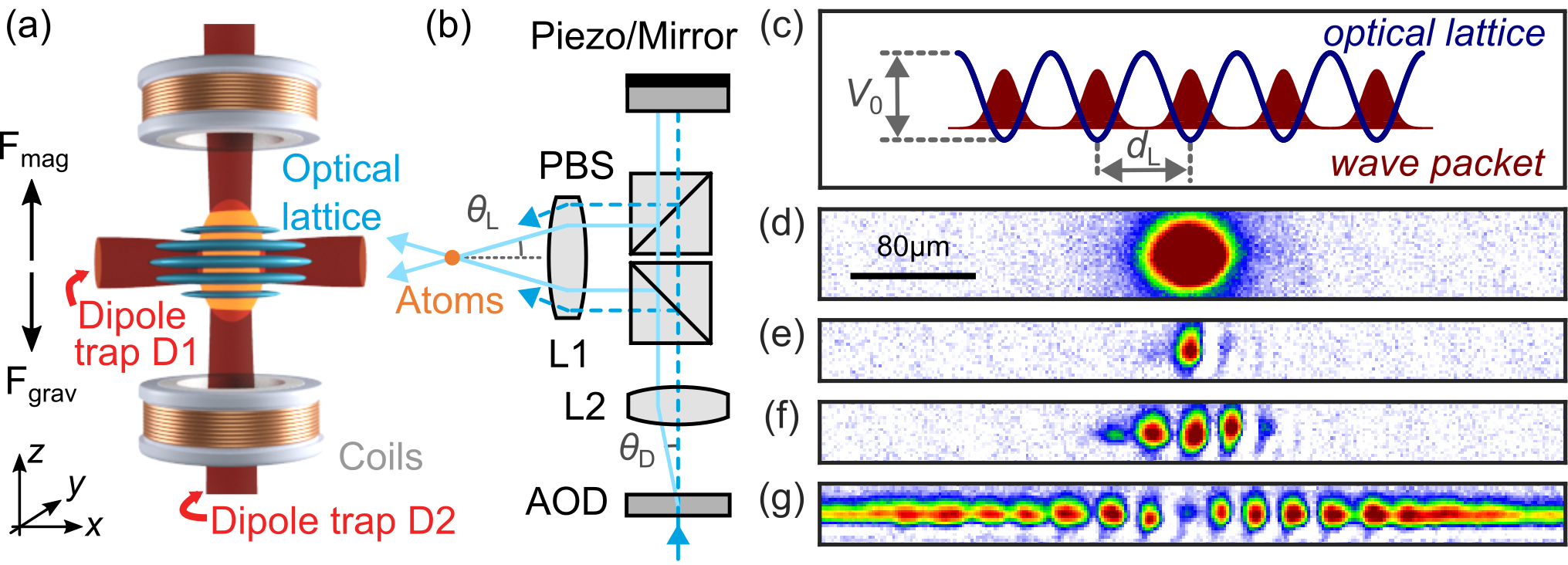}
\caption{Experimental setup and preparation schemes. (a) Configuration of laser beams and coils, and (b) of the accordion lattice. A beam is diffracted by an acousto-optic deflector (AOD) and split into two parallel beams by polarizing beam splitters (PBS). These beams are focused by a lens (L1) onto the atomic cloud and interfere creating the accordion lattice potential. (c) Schematic of the optical lattice and occupied lattice sites. (d-g) \emph{In situ} absorption images of the initially prepared states. (d) Full cloud in approximately twenty lattice sites with $\dL=3\,\upmu$m, which is below the optical resolution limit. (e) Preparation of a wave packet in one lattice site by selectively removing atoms with microwave sweeps, followed by an increase of the lattice spacing to $\dL=27(1)\,\upmu$m. Preparation for (f) a multi-site soliton in 3-5 lattice sites, and for (g) a gray soliton with a single unoccupied site.} \label{Fig:preparation}
\end{figure}

In this article, we present our preparation procedure for the generation of lattice solitons. Our approach utilizes an accordion lattice with tunable spacing $\dL$ [Fig.\,\ref{Fig:preparation}(a) and (b)]~\cite{fallani2005,alassam2010,ville2017}, which allows us to prepare an initial state with a well-defined number of occupied lattice sites and densities, and to magnify the wavefunction for an improved detection of the soliton's density distribution. Accordion lattices provide a versatile tool for cold-atom experiments in preparation and imaging schemes~\cite{fallani2005, williams2008, li2008, peil2003}. They operate by varying the angle $\theta_\text{L}$ between two interfering laser beams, which generate the lattice potential through a periodic intensity pattern. Several methods for controlling $\theta_\text{L}$ have been demonstrated, including adjusting the relative angle between two mirrors~\cite{fallani2005,huckans2009}, changing the beam separation with a translating stage~\cite{li2008,ville2017}, and employing an acousto-optic deflector (AOD)~\cite{williams2008,alassam2010}. In our setup, we combine the mechanically stable beam-splitting setup of Ref.\,\cite{ville2017} with the fast angle control provided by an AOD [Fig.\,\ref{Fig:preparation}(b)]. We achieve a tunable lattice spacing $\dL=1.4(2)\,\upmu$m to $43.1(4)\,\upmu$m, allowing access both to Bose–Hubbard physics with short tunneling times and to individual potential wells of atoms with long tunneling times, suitable for resolving individual lattice sites.

The key steps and variations of our preparation procedure are illustrated in Fig.\,\ref{Fig:preparation}(d)-(g). We begin with a Bose-Einstein condensate loaded into a 1D optical lattice with a spacing $\dL=3\,\upmu$m that is below the resolution limit of our absorption images [Fig.\,\ref{Fig:preparation}(d)]. Spatially selective microwave transitions in combination with resonant laser light are then used to prepare atoms in a well-defined number of lattice sites. To create single-site solitons, we confine the atoms to a single site [Fig.\,\ref{Fig:preparation}(e)], whereas for multi-site solitons we typically prepare them on three to five sites [Fig.\,\ref{Fig:preparation}(f)]. Gray solitons, which are moving dark solitons with reduced density dip \cite{busch2000}, can also be prepared by selectively removing atoms from a single site [Fig.\,\ref{Fig:preparation}(g)]. Finally, we quench the interaction strength and remove the axial confining potential to create attractive lattice solitons.

This paper is laid out as follows: We introduce our experimental setup, the accordion lattice and lattice stabilization technique in Sec.\,\ref{sec:experimentalSetup}. To image individual lattice sites we make use of a magnification scheme, which we introduce and quantify in Sec.\,\ref{sec:excitations}. We discuss and compare various calibration techniques to precisely know the lattice spacing and depth in Sec.\,\ref{sec:accordionCalib}, before quantifying lattice loading in Sec.\,\ref{sec:atomNum}. To generate a singular occupied site in the accordion lattice we used a sequence of resonant microwave pulses, this is discussed in Sec.\,\ref{sec:MWRemoval} before analyzing the efficiency. Finally, we generate solitons in the accordion lattice by quenching the interaction strength and analyze the effect of different quench parameters in Sec.\,\ref{sec:quench}.

\section{Experimental setup}\label{sec:experimentalSetup}

Our experimental sequence consisted of two phases. First, an initial cooling phase was used to create a Bose-Einstein condensate (BEC) of $10^5$ cesium atoms in the $\ket{F=3,m_F=3}$ state~\cite{dicarli2019}. The atoms were trapped by two crossed laser beams with final trap frequencies $\omega_{x,y,z} =2\pi\times(26,32,18)\,$Hz and magnetically levitated against gravity using a field gradient of $\partial_z B = 31\,\text{G}/\text{cm}$ \cite{gustavsson2008,kraemer2004}. A magnetic offset field was applied to control the atomic interaction strength via a broad magnetic Feshbach resonance, with a zero crossing at $17.1$\,G \cite{gustavsson2008,berninger2013}. The second phase of the sequence focuses on the preparation and creation of lattice solitons with the accordion lattice as the central tool.

The working principle of the accordion lattice is illustrated in Fig.\,\ref{Fig:preparation}(b). An incident laser beam is diffracted by an AOD and split into two parallel beams by polarizing beam splitters (PBS) \cite{ville2017}. These beams are focused by a lens (L1) onto the atomic cloud and interfere creating the accordion lattice potential. The lattice spacing at the position of the atoms is given by \begin{equation}\label{eq:latticeSpacing}
    \dL = \frac{\lambda}{2} \frac{1}{\sin \thetaL }, \quad \text{with} \quad \tan \thetaL = \tan \theta_\text{D}\frac{f_\text{L2}}{f_\text{L1}},
\end{equation}
where $\lambda=780\,\text{nm}$ is the blue detuned laser wavelength and $\thetaL$ is the angle between the interfering beams. This angle $\thetaL$ is determined by the focal lengths of lenses L1 ($f_\text{L1}=100$\,mm) and L2 ($f_\text{L2}=500$\,mm), and the lateral displacement of their input beams, which again depends on the deflection angle $\thetaD$ and $f_\text{L2}$. The AOD provides a bandwidth of $20$\,kHz, enabling precise and dynamic control of $\dL$ and allowing for the generation of time-averaged lattice potentials. The lattice position is actively monitored via a separate interference pattern imaged on a CCD camera. Feedback stabilization is achieved with a piezo-mechanical actuator that adjusts the position of a mirror, thereby controlling the lattice phase $\varphi_\text{L}$.

\section{Detection with magnification}\label{sec:excitations}
\begin{figure}
\includegraphics[width=\columnwidth]{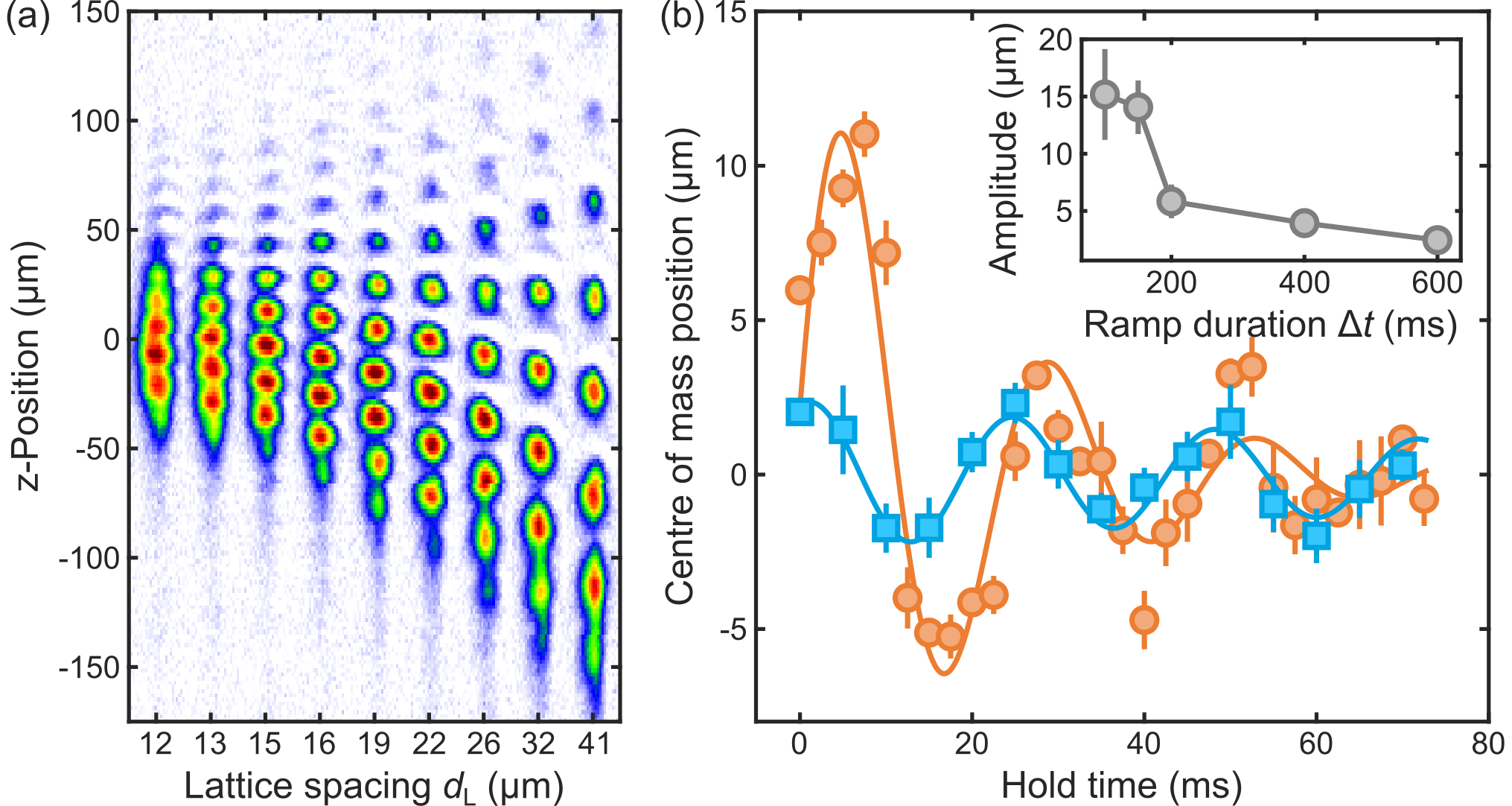}
\caption{Magnification of density profile. (a) Absorption images of density profile in the optical lattice after a short expansion time of $2$\,ms. Parameters $V_0>200\,\Er$, ramp $\Delta t=400\,$ms. Shift of center position of $0.87\,\dL$. (b) Oscillation of wave packet in central lattice site after magnification to $\dL=30\,\upmu$m in $150$\,ms (orange circles) and $600\,$ms (blue squares). Solid lines denote damped sine fits to the data. Inset: measured oscillation amplitude for varying ramp duration $\Delta t$, error bars denote the $2\sigma$ error, line is a guide to the eye.}\label{Fig:magnification}
\end{figure}

One of the primary roles of the accordion lattice in our sequence is to increase the lattice spacing at the end of the experimental cycle, 'magnifying' the sites to resolve them individually~\cite{alassam2010,cruickshank2025}. In our setup, individual sites are distinguishable for $\dL\ge6\,\upmu$m. However, we magnified the lattice to approximately $\dL=20\,\upmu$m to provide a clear separation of sites for measurements of atom number.

To magnify the atomic density distribution we suppressed tunneling by increasing the lattice depth to $V_0=100\,\Er$ in $5$\,ms, followed by a slower increase of $\dL$ to $20\,\upmu$m in $400$\,ms. The final lattice depth was greater than $300\,\Er$ with a tunneling time of $h/J>10\,$s, where $J$ is the tunneling matrix element for atoms in the lowest band. In addition to an improved resolution, this magnification sequence reduces the atomic density in each lattice site, avoiding the re-absorption of imaging light which could produce an apparent increase in atom number~\cite{veyron2024}.

The magnification inherently shifts off-center lattice sites due to an increase in the spacing $\dL$. Additionally, we observed a small shift of the cloud position [Fig.\,\ref{Fig:magnification}(a)] caused by variations in the intersection point of the lattice beams as the angle $\theta_\text{L}$ was adjusted. Both displacements can induce center-of-mass oscillations of the atoms within each site that persist even after the magnification process. However, the measurements of the final occupation numbers per site were unaffected by those oscillations, and we ignored them except for the calibration of $\dL$. The amplitude of those oscillations was reduced by increasing the duration of the magnification ramp (see below). A second challenge arises from the finite size of the lattice beams with a waist of $200\,\upmu$m.  Depending on the magnification, this finite size limited the number of observable lattice sites, e.g., to approximately ten sites for a final spacing of $\dL = 20\,\upmu$m. Due to the Gaussian intensity profile of the lattice beams, the lattice depth decreased at larger distances from the center, allowing atoms in outer sites to escape.

Absorption images in Fig.\,\ref{Fig:magnification}(a) illustrate the density distribution for increasing $\dL$ after magnification. The images reveal the shift of the lattice center position (here $0.87\,\dL$) as well as the loss of atoms at the edge of the lattice beams. For the study of solitons, we adjusted the magnification to the specific problem, using a larger magnification for single-site solitons compared to multi-site solitons. The amplitude of the center-of-mass oscillations depends strongly on the duration of the magnification ramp $\Delta t$ [Fig.\,\ref{Fig:magnification}(b)]. For $\Delta t = 150\,$ms, the oscillations show an initial amplitude of $14.2(2.5)\,\upmu$m ($0.46\,\dL$), which was significantly reduced to $2.4(0.7)\,\upmu$m for a slower ramp of $\Delta t = 600\,$ms. To balance experimental run time and minimize excitations, we chose an intermediate value of $\Delta t = 400\,$ms [see inset of Fig.\,\ref{Fig:magnification}(b)].

\section{Calibration of the accordion lattice}\label{sec:accordionCalib}

Our lattice setup enables dynamic control of the lattice spacing $\dL$ over a wide range, from $1.4(2)\,\upmu$m to $43.1(4)\,\upmu$m, allowing us to explore the system in three regimes. In the tight-binding limit, characterized by deep lattice potentials and small lattice spacing, the system exhibits a well-defined band structure with its characteristic resonances. In contrast, in the shallow lattice regime, the periodic potential acts as a perturbation to the underlying continuous medium. While lattice solitons can emerge in both regimes \cite{maluckov2008}, our focus here is on shallow lattices. We use atom numbers $N$ between \num{200} and \num{3000} and lattice depth $V_0\leq6\,\Er$, where $\Er=\hhbar \kL/(2m)$ is the recoil energy and $\kL=\pi/\dL$ the lattice momentum, and $m$ the atomic mass. The parameters were chosen to study the system with low three-body loss rates and on experimentally convenient times scales below one second. The third regime arises during the magnification process, where we suppress tunneling in a deep lattice with large spacing, and the system can be effectively described as an array of isolated harmonic traps.

\subsection{Calibration of the lattice spacing}\label{sec:latticeSpacing}

\begin{figure}
\includegraphics[width=\columnwidth]{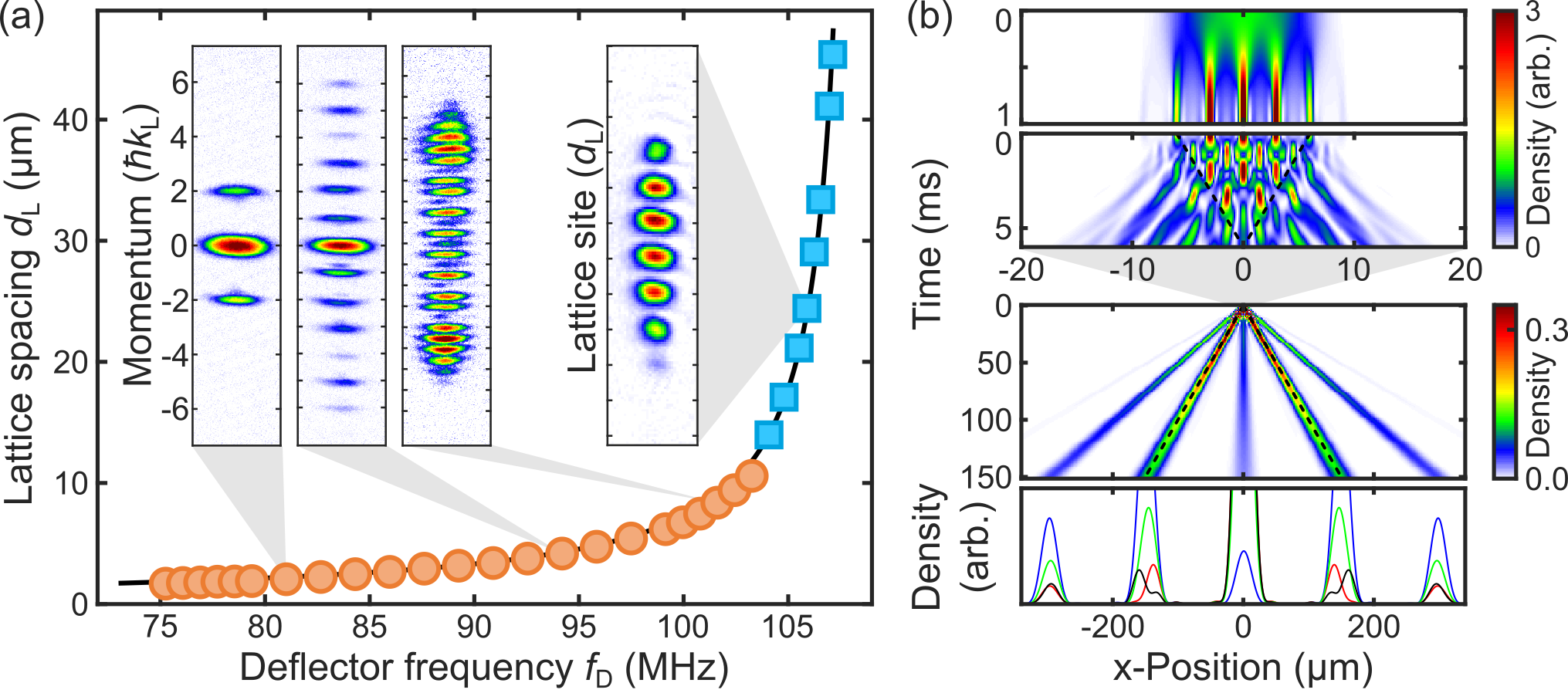}
\caption{Calibration of lattice spacing $\dL$. (a) Measurement of lattice spacing by determining the atom position (blue squares) and the momentum after expansion from the lattice (orange circles). Insets show absorption images after an expansion time 122\,ms for deflector frequency $f_\text{D}=82.7$\,MHz, 94.2\,MHz, 101.6\,MHz, and in position space for 104.9\,MHz (left to right). (b) Numerical simulation of the time evolution shows the density distribution for parameters $V_0=20\,\Er$, $\dL=3\,\upmu$m, $a_\text{s}=10\,a_0$, $N=10,000$ atoms. Panels, from top to bottom, show the evolution during 1\,ms lattice pulse, the initial evolution with Talbot revivals, and the full evolution over 150\,ms. Black dotted lines indicate the $\pm 2\hhbar\kL$ momentum. Bottom: The final momentum peaks of the density distribution show a small dependence on pulse duration: 1.00\,ms (blue line), 1.45\,ms (green line), 1.70\,ms (red line), and 2.00\,ms (black line).\label{Fig:latticeCalibration2}}
\end{figure}

Due to its variability, accurately calibrating the lattice spacing requires two experimental methods, each suited to different regimes of $\dL$ \cite{fallani2005}. For $\dL>10\,\upmu$m, we directly resolved individual lattice sites by imaging the atomic density distribution after a time-of-flight expansion of $2\,$ms, during which we switch off the magnetic field [rightmost inset in Fig.~\ref{Fig:latticeCalibration2}(a)]. We extracted the center positions of the density peaks using Gaussian fits and determined $\dL$ by averaging the distance between neighboring peaks [blue squares in Fig.~\ref{Fig:latticeCalibration2}(a)]. The image pixel size was calibrated independently through comparison with the gravitational acceleration of the BEC.

For smaller spacings, we applied a short pulse of the lattice potential and determined the lattice momentum $\hbar\kL$ from the atoms’ momentum distribution after $122\,$ms of free expansion [three leftmost insets in Fig.~\ref{Fig:latticeCalibration2}(a)]. This method, introduced in Refs.\,\cite{ovchinnikov1999,denschlag2002}, relies on the fact that the momentum profile after release from the lattice is dominated by discrete peaks at multiples of $2\hbar \kL$, which are visible in the absorption images. We extracted the center momentum of each peak and calculated $\kL$ from the spacing between the peaks. The lattice spacing was obtained using the relation $\kL=\pi/\dL$ [orange circles in Fig.~\ref{Fig:latticeCalibration2}(a)]. A variation of this method, where the atoms were adiabatically transferred into the lattice rather than pulsing the lattice potential, proved less suitable due to the loss of coherence at large lattice spacings.

The two methods diverge by approximately $10\%$ around $\dL = 10.5\,\upmu$m [Fig.\,\ref{Fig:latticeCalibration2}(a)] (see below). We expect the direct spatial imaging to be accurate within the limits of our optical resolution and our ability to determine the center position at each site using a Gaussian fit. In contrast, the momentum-based measurement increasingly deviates and becomes unreliable for larger lattice spacings, as discussed below. To perform a complete calibration of $\dL$, we excluded data points in the range $\dL=6-10.5\,\upmu$m, and fit the remaining data using
\begin{equation}\label{eq:latticeSpacingFit}
    \dL = A\frac{\sqrt{1+B^2(f_\text{D}-C)^2}}{B(f_\text{D}-C)},
\end{equation}
which is a reformulation of Eq.\,(\ref{eq:latticeSpacing}) for small angles $\theta_\text{D}$ and deflector frequency $f_\text{D}$ with fitting parameters $A$, $B$, and $C$ [black line in Fig.\,\ref{Fig:latticeCalibration2}(a)]. We found the lower limit of $\dL$ to be set by the maximum angle $\theta_\text{L}$ permitted by the optical setup, while the upper limit is determined by the minimum achievable distance between the parallel beams produced by the beam splitters, typically constrained by the beam size [Fig.\,\ref{Fig:preparation}(b)].

To understand the small mismatch between the two methods we use the 1D-Gross-Pitaevskii equation to simulate the time evolution of the atoms' density profile when pulsing the lattice potential with variable duration $t_\text{p}$ \cite{bao2003} [Fig.\,\ref{Fig:latticeCalibration2}(b)]
\[
i\hbar \frac{\partial \psi(z,t)}{\partial t}
=
\left[
-\frac{\hbar^{2}}{2m}\frac{\partial^{2}}{\partial z^{2}}
+ V(z,t)
+ g_{1\mathrm{D}}\, |\psi(z,t)|^{2}
\right]
\psi(z,t).
\]

Here, $V(z,t)$ is the combined confining potential of lattice and trapping beams D1 and D2,  $g_{1\mathrm{D}} = 2\,\hbar \omega_{\perp}\, a_\text{s} N_\text{tot}$ is the effective coupling constant for $N_\text{tot}$ atoms with radial trap frequency $\omega_\perp$, and $n(z,t)=|\psi(z,t)|^2$. First, the wave packet's ground state in the crossed dipole trap was determined with imaginary time propagation~\cite{chiofalo2000}. Second, a split-step Fourier method was used to propagate the wave function in time~\cite{suarez2013}. Top to bottom panels in Fig.\,\ref{Fig:latticeCalibration2}(b) show the atoms' time evolution during the different steps of the simulation. During the pulse duration, the wave packet oscillates between different lattice bands~\cite{ovchinnikov1999,denschlag2002} with the density distribution reflecting the states in the bands. After the pulse follows a brief time interval during which a Talbot-like interference pattern develops~\cite{santra2017}. This pattern results from the interference of matter waves expanding from different lattice sites. It starts to fade once the wavefronts with the dominant momentum components $\pm 2\hbar \kL$ from the outermost sites reach the central site (black dashed lines). Finally, after $150$\,ms of free expansion a clear pattern of momentum components is visible. We used the peak positions of this pattern to determine $\dL$, assuming a linear expansion with peaks at multiples of $h/(m\dL)t$. The simulation shows good agreement with our measurements and with previous numerical results \cite{fioroni2024}. However, we relied on it primarily to understand the mismatch in our calibrations, as discussed below.

Determining $\dL$ by pulsing the lattice potential has three potential limitations at large values of $\dL$. First, resolving the $2\hhbar \kL$ momentum components after expansion can become increasingly difficult, as the distances between the peaks scale with $1/\dL$. However, this is not a limiting factor in our case due to long expansion times at negligible interaction strength. Second, coherence between lattice sites can decrease at large lattice spacings, leading to a diminished interference pattern. For our measurements, this effect is negligible due to the short pulse durations. However, the loss of coherence becomes a significant limitation when measuring the free expansion after adiabatically loading into a lattice. Finally third, the simulation reveals that the peak position can shift depending on the pulse duration [bottom panel in Fig.\,\ref{Fig:latticeCalibration2}(b)]. The momentum of the peaks varies slightly as the atoms oscillate between bands during the pulse duration. We believe this last effect accounts for the discrepancy observed between the two calibration methods in Fig.\,\ref{Fig:latticeCalibration2}(a).

\subsection{Calibration of the lattice depth}\label{sec:depthCalib}

\begin{figure} [t]
\includegraphics[width=\columnwidth]{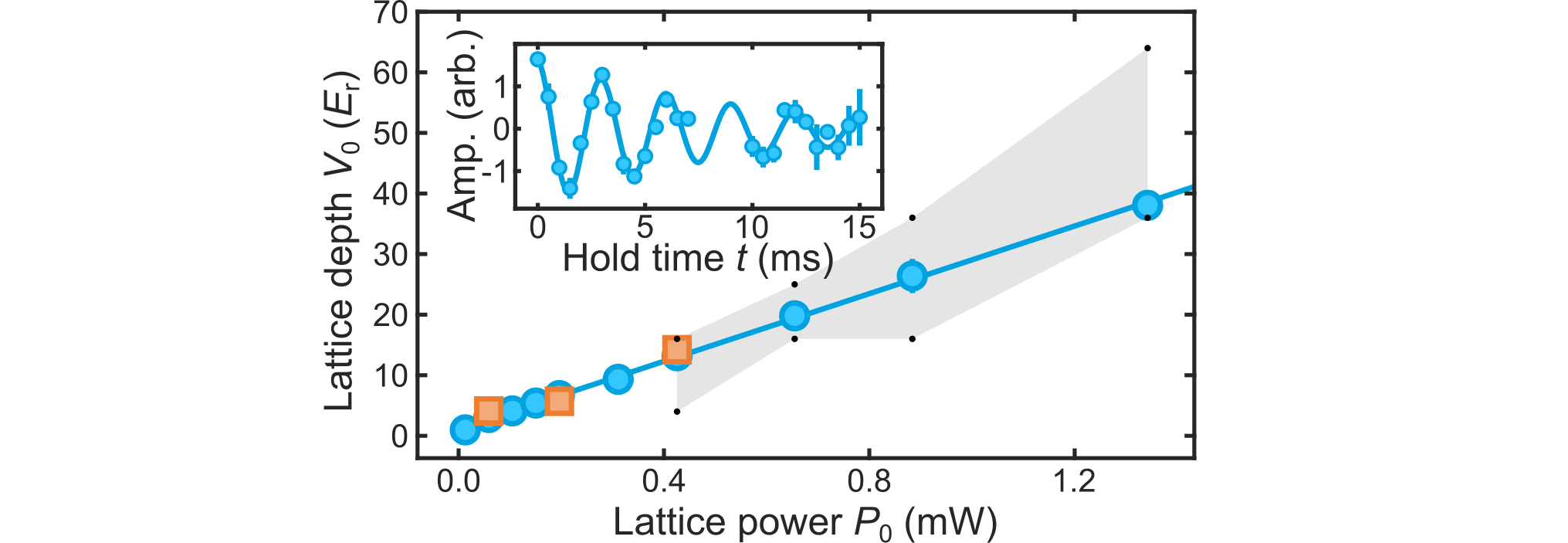}
\caption{Calibration of lattice depth $V_0$. Determination of $V_0$ by measuring the trap frequency $\omega_\text{L}$ at each lattice site via center-of-mass oscillations (blue circles), and by measuring oscillations between energy bands (orange squares). The gray shaded area indicates the range of depths achievable using the method described in Ref.\,\cite{huckans2009}, with black dots showing the data points ($\dL=3\,\upmu$m throughout). Inset shows measured oscillation for $P_0=0.75\,$mW, solid line is a damped sinusoidal fit.}\label{Fig:depthCalibration}
\end{figure}

Similar to the calibration of $\dL$ in Sec.\,\ref{sec:latticeSpacing}, determining the lattice depth $V_0$ also requires multiple methods to cover the full range of lattice spacings. We determined $V_0$ in units of the recoil energy, which depends on the laser power and lattice spacing, and calibrated it across the full parameter space spanned by $P_0$ and $\dL$. For clarity, however, we present here only a representative slice of this space at fixed lattice spacing $\dL = 3.0\,\upmu$m.

We first measured $V_0$ by measuring the trap frequency $\omega_\text{L}$ at individual sites by directly monitoring the oscillation of atoms within a single well. To prepare the system, we loaded approximately $2,000$ atoms into the central lattice site using a tightly confining dipole trap (beam D1) with an axial trapping frequency of $\omega_z = 2\pi \times 100\,\text{Hz}$. We then slowly shifted the phase $\varphi_\text{L}$ of the optical lattice by adjusting a piezo-driven mirror [Fig.\,\ref{Fig:preparation}(b)] over 100\,ms, and simultaneously removed the potential of D1. Oscillations were initiated by rapidly returning $\varphi_\text{L}$ to its original value within 2\,ms. The resulting oscillations of the atoms were measured after an expansion time of 22\,ms [inset in Fig.\,\ref{Fig:depthCalibration}]. To ensure that the atoms remained within a single site, we adjusted the magnitude of the phase shift for different lattice depths. The lattice depth in $\Er$ was calculated using
\begin{equation} 
\frac{V_0}{\Er} = \frac{\omega_\text{L}^2 m^2}{\hbar^2 \kL^4} \text{\;and\;} V_0 = C P_0,  \label{eq:depthCalib}
\end{equation}
where $C$ is a constant that contains beam parameters, such as waist, overlap, and frequency detuning from atomic transitions. The resulting measurements show the expected linear dependence of lattice depth on beam power $P_0$ [blue circles in Fig.\,\ref{Fig:depthCalibration}]. This method is limited to sufficiently large values of $P_0$ and $\dL$, such that the lattice trap frequency remains larger than the frequency of inter-site tunneling.

The second method for calibrating $V_0$ was based on interband oscillations of the atoms \cite{ovchinnikov1999,denschlag2002,morsch2006} and was better suited for lattices with $\dL\le3\,\upmu$m as there is a well defined band structure. In this approach, the lattice was rapidly pulsed on for a short duration $t_\text{p}$, placing the atoms in a superposition of states in the odd energy bands. Band populations oscillate during $t_\text{p}$ with frequencies that depend on the energy gaps between the bands \cite{ovchinnikov1999}. We determined the band population in quasi-momentum space using a band-mapping technique and a free expansion time of $120\,$ms. Our measurements [orange squares in Fig.\,\ref{Fig:depthCalibration}] are in good agreement with the values obtained from the position-space oscillation method. The method works usually best for shallow lattices when only one energy band is excited. 

Finally, the last method to calibrate $V_0$ employed a variation of the previous technique \cite{huckans2009}. The lattice is again imposed onto the atoms for only a short period of time ($t_\text{p}\sim1\,$ms) until the momentum distribution shows a maximum number of modes $N_\text{max}$. Assuming that the momentum modes are created by atoms in odd bands, the lattice depth can then be approximated using $V_0/\Er=4\,N_\text{max}^2$. This method works best for deep lattices when many bands are excited. However, the discreteness of $N_\text{max}$ presents a challenge for shallow lattices due to the a large relative uncertainty of $N_\text{max}$. We used a shaded area in Fig.\,\ref{Fig:depthCalibration} to indicate this uncertainty, assuming $\Delta N_\text{max}=\pm 1$. Though less precise, the results agree well with the previous measurements.

\section{Initial state preparation}\label{sec:StatePreparation}
\begin{figure}
\includegraphics[width=\columnwidth]{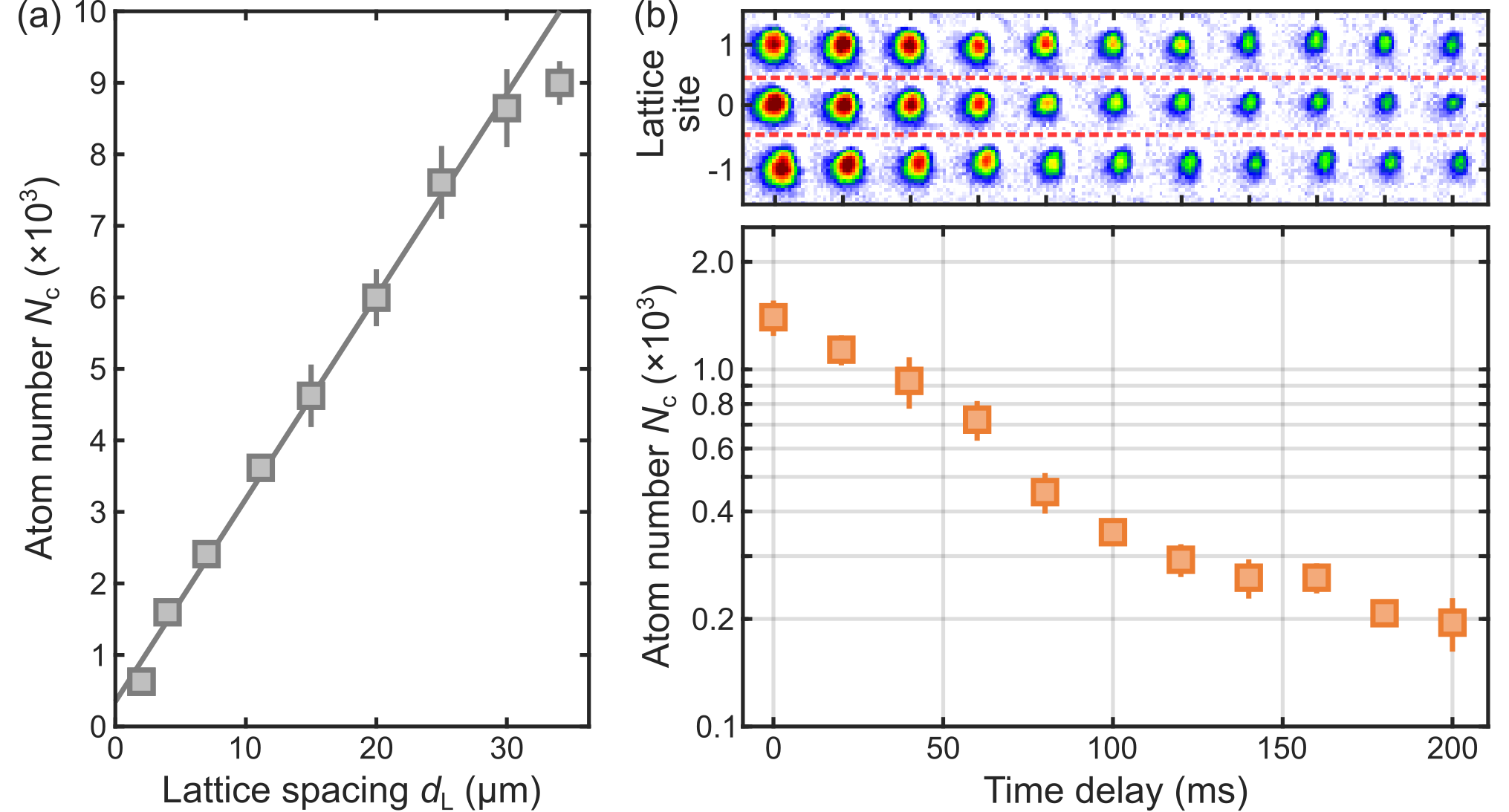}
\caption{Atom number preparation in central lattice site. (a) Atom number reduction by loading into a lattice with spacing $\dL$ and depth $V_0\approx30\,\Er$ (gray squares), where $\Er$ is defined at $\dL=3\,\upmu$m. $N_\text{tot}=13.0(4)\times10^4$ atoms, error bars denote the standard deviation over five repetitions, solid line is a linear fit to the data. (b) Atom number in central site after a time delay between switching off the dipole trap D1 and adding the accordion lattice. Time $t=0$\,ms is the moment immediately after the dipole trap is removed. (Top panel) Absorption images of corresponding density profile of central three lattice sites, dashed red lines delineate center site. (lower panel) Extracted atom number in the central site, error bars denote the standard deviation over seven repetitions.}\label{Fig:atomNumberControl}
\end{figure}

Lattice solitons show a strong dependence on atom number and as such controlling the occupation of individual sites is a central challenge for the preparation of the initial state. An advantage of using an accordion lattice is the ability to dynamically control this occupation during loading, not only by altering the lattice depth but also the lattice spacing. We employed a two-step process for the preparation of the density distribution, first by preparing a small sample of $N\lesssim30,000$ in the lattice potential, and second, by removing atoms in unwanted sites with a combination of a microwave transfer and a resonant laser beam. We adjusted the trap frequency of dipole trap D1 during the process to keep the density distribution close to the system's ground state, and provided a final waiting period to allow the system to settle.

\subsection{Transfer of atoms to the accordion lattice}\label{sec:atomNum}
We found it essential to prepare the desired atom number in the central lattice site, $N_\text{c}$, already during the lattice loading procedure, thereby reducing excitations during the subsequent atom removal process and quench. To prepare the sample in the lattice, we first reduced the atom number of the Bose-Einstein condensate in the crossed dipole trap from 130,000 to 30,000 by reducing our magnetic levitation gradient for 3\,s. We then increased the lattice depth $V_0$ over $150\,$ms to a final value while simultaneously adjusting the trapping frequency $\omega_z$ of the dipole trap to create the intended density distribution. The final atom number per site is controlled by $\omega_z$, $a_\text{s}$, $V_0$, and $\dL$. Especially the reduction $\dL$ allows us to vary $N_\text{c}$ by a factor of ten [Fig.\,\ref{Fig:atomNumberControl}(a)]. The approach works well down to approximately 1000 atoms per site.

To further reduce $N_\text{c}$ to a few hundred atoms, we lowered the depth of the dipole potential to zero in $250\,$ms and added a time delay, allowing the atoms to slowly expand in the vertical dipole potential of beam D2 before increasing $V_0$ in $150\,$ms. Fine control of the atom number was possible by increasing the time delay between these two ramps, allowing the atoms to expand further before their motion was frozen by the lattice potential. The absorption images in Fig.\,\ref{Fig:atomNumberControl}(b) demonstrate this reduction for the central three sites, while the lower plot shows the extracted atom number in the central site. 

\subsection{Microwave-assisted atom removal}\label{sec:MWRemoval}

We employed a combination of microwave transitions and short pulses of resonant light to prepare a single occupied lattice site, which served as the starting point for soliton creation described in the next section. Microwave-assisted atom removal is a common technique for creating single occupied layers in quantum gas microscopes~\cite{Sherson2010, weitenberg2011, haller2015, peaudecerf2019}. Addressing individual lattice sites is challenging at small lattice spacings of only a few hundred nanometers, as it requires precise control of magnetic field gradients and lattice angles~\cite{peaudecerf2019}. Accordion lattices simplify this preparation process due to their variable lattice spacing. We studied the preparation process in two steps: we first optimized the removal of atoms in a single site of the accordion lattice before implementing the removal of atoms in all sites but one.

We use the levitating magnetic field gradient, $\partial_z B$ (see Sec.\,\ref{sec:experimentalSetup}), to enable selective spatial addressing of atoms along the $z$-direction. The gradient shifts the energy levels of the Zeeman substates $\ket{F=3,m_F=3}$ and $\ket{F=4,m_F=4}$ that were used for the microwave transition. The gradient makes the microwave transitions spatially dependent with a frequency shift of $7.5\,\text{kHz}/\upmu\text{m}$. To achieve site-selective transfer, we employ hyperbolic secant (HS1) pulses~\cite{Khudaverdyan2005}, which maximize population transfer at resonant lattice sites while suppressing transfer in neighboring sites. These pulses produce a flat-top transfer profile with sharp spectral edges~\cite{garwood2001, peaudecerf2019} by sweeping the microwave detuning $\delta(t)$ and coupling amplitude $\Omega(t)$ according to hyperbolic tangent and secant functions, respectively
\begin{equation}\label{eq:MWpulses}
    \delta(t)=\left(\delta_0/2\right)\text{tanh}(2t/\tau),\,
    \Omega(t)=\Omega_0\,\text{sech}(2t/\tau),
\end{equation}
where $\Omega_0=9.9(1.0)\,$kHz is the Rabi frequency of the transition, $\delta_0$ is the frequency width of the pulse and $\tau$ is a characteristic timescale. The smoothness of the switch on was controlled using the ratio $\tau/T=1/3$, where $T$ is the pulse duration. After transfer, atoms in the $\ket{F^\prime=4}$ state are no longer levitated and accelerate downwards. To prevent collisions with other atoms, we remove them using a $1\,$ms pulse of a horizontal laser beam resonant with the $\ket{F=4 \rightarrow F^\prime=5}$ transition~\cite{kuhr2005}.

\begin{figure}
\includegraphics[width=\columnwidth]{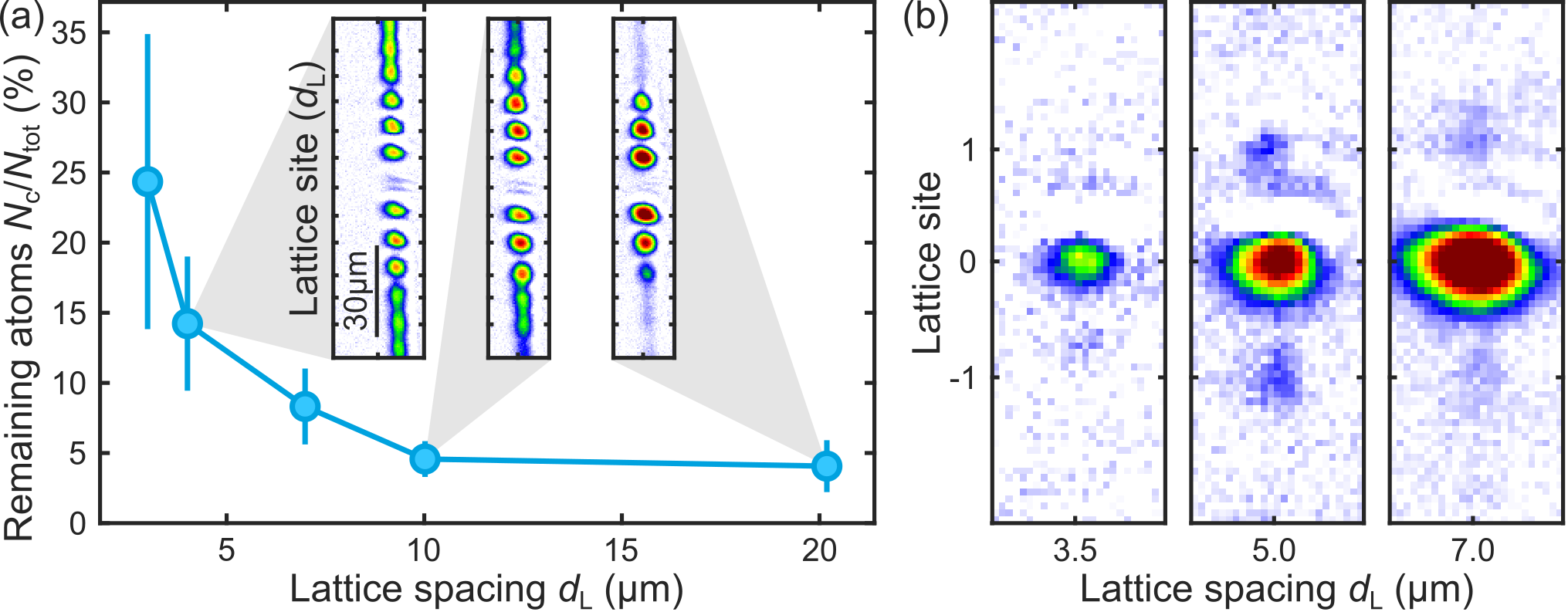}
\caption{Microwave-assisted atom removal. (a) Fraction of atoms remaining in the central site after the removal process for varying $\dL$. Sweep parameters of the HS1 pulse were optimized for each lattice spacing with widths of $\delta_0=15\,$kHz and $90\,$kHz for $\dL=3\,\upmu$m and $20\,\upmu$m, respectively (lattice depth was $V_0=216\,\Er$ defined at $\dL=3\,\upmu$m throughout). The insets show averaged absorption images for different $\dL$. (b) Averaged images of single-site preparation in lattices with spacing $\dL=3.5\,\upmu$m, 5$\,\upmu$m, and $7\,\upmu$m. Images show density profile after magnification.}\label{Fig:MWRemoval}
\end{figure}

We first characterized the removal efficiency of our scheme for atoms in the central lattice sites [Fig.\,\ref{Fig:MWRemoval}(a)]. For each lattice spacing $\dL$, we optimized the HS1 pulse parameters to minimize the number of atoms remaining in the target site. Depending on $\dL$, the frequency sweep widths ranged from $\delta_0=15$ to $90$\,kHz. After all sweeps, the neighboring sites remained unaffected, allowing us to apply the pulses more than once to achieve larger efficiencies. However, for a single pulse in lattices with spacings larger than $10\,\upmu$m, approximately $5\%$ of the atoms remained, whereas for $\dL=3\,\upmu$m this fraction increased to $25\%$. The remaining atoms show as two weak horizontal stripes in the \emph{in situ} images [insets Fig.\,\ref{Fig:MWRemoval}(a)], which is an artifact of our imaging system caused by diffraction from small objects. 

In a second measurement, we studied the preparation of only one occupied site. Our scheme used two broad HS1 pulses with frequency sweeps that addressed atoms above and below the site. Figure~\ref{Fig:MWRemoval}(b) shows images of the preparation of a single lattice site for varying lattice spacings, where a single site could be reliably prepared in lattices with spacing $\dL=3-3.5\,\upmu$m with minimal atoms in neighboring sites [leftmost image in Fig.\,\ref{Fig:MWRemoval}(b)]. The variation of the atom number between the images is a result of the loading process and was not from the removal of atoms by the microwave scheme. All images in Fig.\,\ref{Fig:MWRemoval}(b) show the preparation after the two frequency sweeps, each of $T=20$\,ms duration, however we increased the sweep width $\delta_0$ by $\sim20\,$kHz when $\dL=7\,\upmu$m to account for the increased lattice spacing.

For the single-site preparation, we achieved a removal efficiency of $90\%$ for atoms in the outer sites, while less than $6\%$ of the atoms in the central site were removed. To verify that the preparation sequence did not induce significant heating of the remaining atoms, we repeated the sequence while leaving two sites occupied and performed expansion-time measurements to observe interference fringes. These fringes persisted even after a $1$\,s hold time in the accordion lattice, confirming the absence of any significant excitations.

\section{Creation of lattice solitons}\label{sec:quench}

Lattice solitons preserve their shape through a balance of kinetic, interaction, and potential energies. Their shape remains stable because density variations increase the system’s energy, with large deviations suppressed by energy barriers. Remarkably, in the presence of a lattice potential, this stabilization mechanism applies not only to attractive but also to repulsive interactions. These wave packets, sometimes called “repulsive solitons” \cite{trombettoni2001}, are related to repulsively bound pairs in the two-atom limit \cite{denschlag2002} and to self-trapping in two-site Josephson junctions \cite{smerzi1997,albiez2005}. We demonstrate in this section that repulsive solitons provide a convenient intermediate stage to create attractive single-site solitons with an interaction quench.

\begin{figure} [b]
\includegraphics[width=\columnwidth]{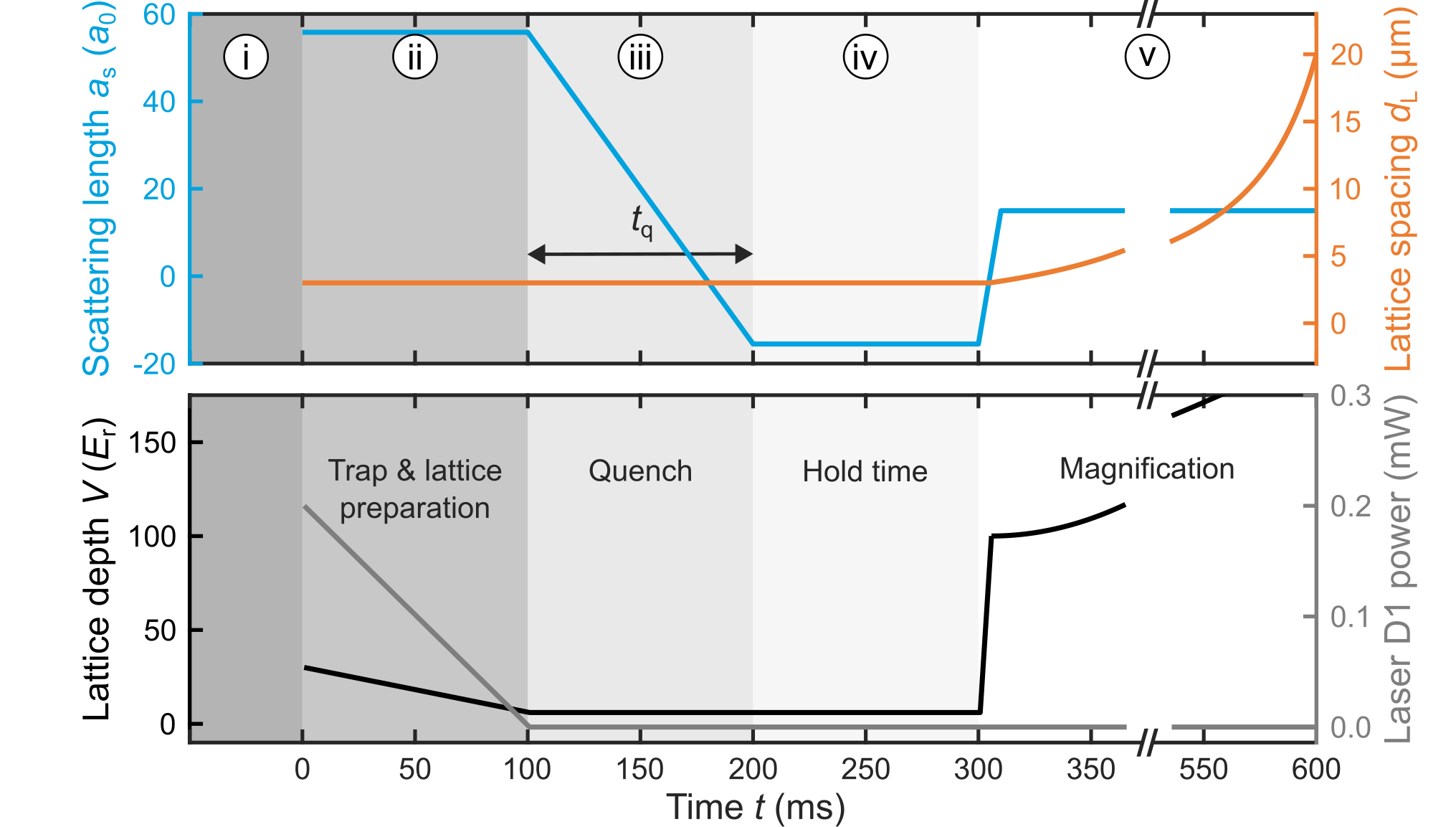}
\caption{Sketch of the experimental sequence for the generation of lattice solitons. Values as a function of time of system variables such as scattering length (blue), lattice spacing (orange), lattice depth (black) and power of horizontal trapping potential D1 (gray). The different experimental phases are marked (i) preparation of lattice site occupations, (ii) slow reduction of trapping potentials, (iii) quench of scattering length, (iv) soliton generation and subsequent hold time, (v) magnification sequence and imaging. The start point $t=0$ occurs immediately after the microwave preparation discussed in Sec.\,\ref{sec:MWRemoval}.}\label{Fig:sequence}
\end{figure}

A common approach to generate solitons in quantum gases involves a quench of both the interaction strength and the confining potential \cite{donley2001,cornish2006,dicarli2019b,cruickshank2025}. Typically, a trapped wave packet is first prepared with repulsive interactions. Then, the confining potential is switched off, and the scattering length is tuned from positive to negative. The quench is designed so that trapping and interaction parameters closely match the initial density distribution, usually a Thomas-Fermi profile, to the density of a stable soliton, often approximated by a sech-squared shape. Strong mismatch in the density profiles can lead to excitations, resulting in atom loss by shedding or to oscillations \cite{dicarli2019b, luo2020}.

Using repulsive solitons in a lattice potential as an intermediate step is convenient, because they can be created by directly reshaping the trapping potential without the possibility of collapse. Our quench protocol to create an attractive single-site soliton is sketched in Fig.\,\ref{Fig:sequence}, with different periods in the experiment designated by Roman numerals. We first prepared a wave packet of $500$ atoms in a single lattice site following the procedure outlined in Sec.\,\ref{sec:StatePreparation}, [region (i) in Fig.\,\ref{Fig:sequence}]. During this stage, the scattering length was fixed at $a_\text{s}=+55\,a_0$, with lattice parameters $\dL=3\,\upmu$m and $V_0=30\,\Er$ [upper and lower plot in Fig.\,\ref{Fig:sequence}]. We then used the stability of the repulsive soliton to slowly approach the target configuration by reducing the lattice depth to $V_0=6\,\Er$ and switching off the external trap over 100\,ms [region (ii)].
Finally, we quenched the system by reducing the scattering length to its final value within a time $t_\text{q}$ [region (iii)] and allowed the soliton to evolve for $100$\,ms [region (iv)]. The density distribution was subsequently measured using absorption imaging after a magnification sequence [region (v)], quickly freezing out the occupation number with an increase in lattice depth to $V_0=100\,\Er$ in $5$\,ms and a slow increase of the lattice spacing as described in Sec.\,\ref{sec:experimentalSetup}. To reduce three-body loss and excitations caused by attractive interactions we tuned the scattering length to $+15\,a_0$ during the magnification sequence.

\begin{figure}[t]
\includegraphics[width=\columnwidth]{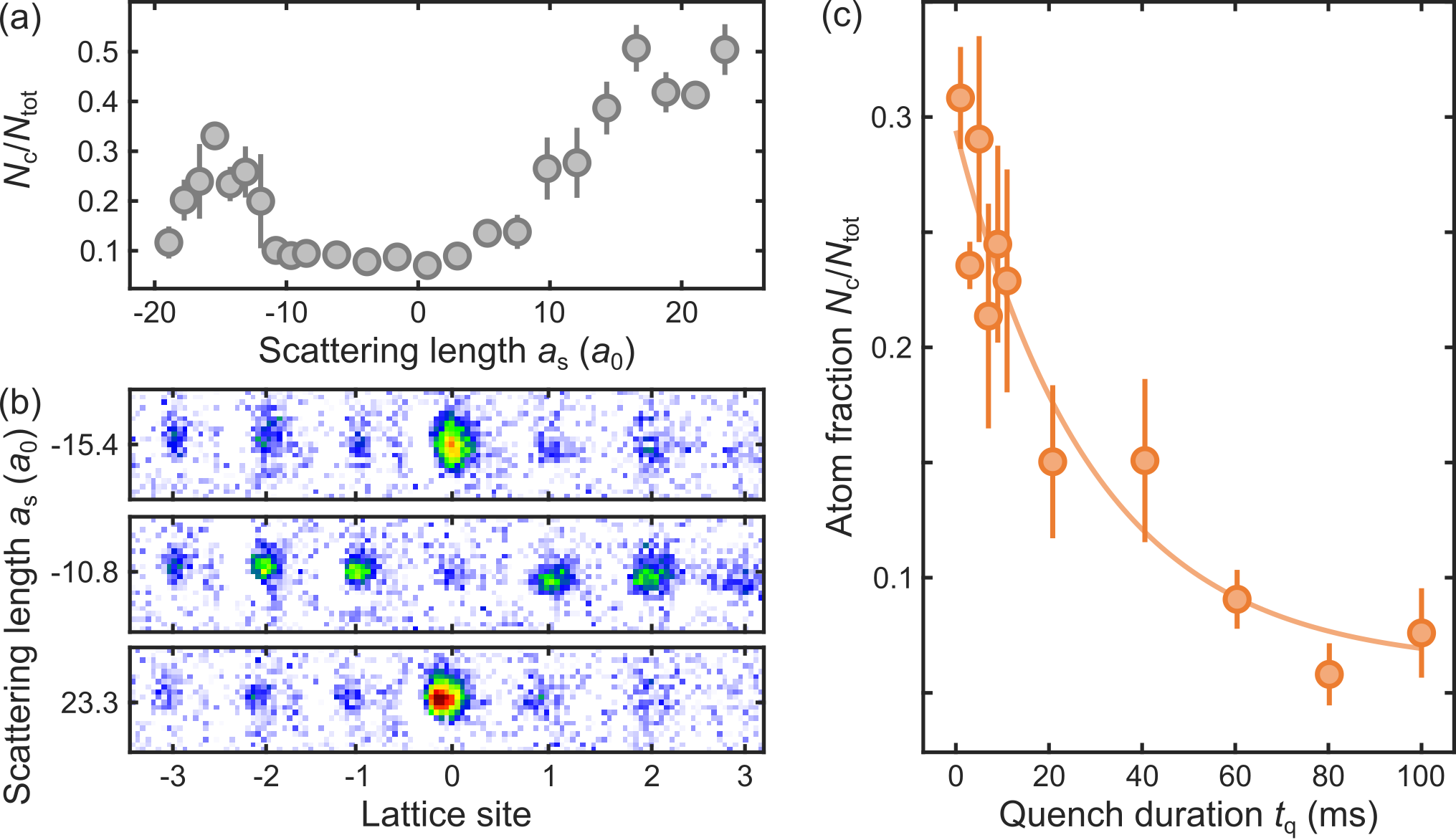}
\caption{Effect of quench duration on soliton stability. (a) Measured fraction of atoms in the centrally prepared site $N_\text{c}/N_\text{tot}$ $100\,$ms after a quench of the scattering length $a_\text{s}$ of varying strength, $t_\text{q}=1\,$ms. (b) Averaged absorption images showing atomic density $100$\,ms after quench to $a_\text{s}=-15\,a_0$ (upper, soliton), $-11\,a_0$ (middle, dispersion) and $+23\,a_0$ (bottom, repulsive soliton). (c) Optimization of the quench duration. $N_\text{c}/N_\text{tot}$ $50\,$ms after the end of the quench with varying duration $t_\text{q}$. The solid line denotes an exponential fit to the data. Faster quenches were more favorable for soliton formation. Parameters were $\dL=3\,\upmu$m, $V_0=6\,\Er$, $a_\text{s}=-15\,a_0$ throughout.}\label{Fig:quench}
\end{figure}

Images in Fig.\,\ref{Fig:quench}(b) show averaged density profiles $100$\,ms after quenching to $a_\text{s}=-15\,a_0$, $-11\,a_0$, and $+23\,a_0$. For final scattering lengths close to zero the wave packet disperses, whereas it remains stable for scattering lengths near $-15\,a_0$. This behavior is also reflected when studying the final relative atom number in the central lattice sites, $N_\text{c}/N_\text{tot}$ which shows a maximum at $-15\,a_0$ due to the formation of attractive solitons [Fig.\,\ref{Fig:quench}(a)]. When quenching to smaller scattering lengths, we exceed the critical interaction strength $g_c$ and the wave packet collapses.

To optimize the quench duration, we fixed the final scattering length to $a_\text{s}=-15\,a_0$ and varied the quench duration $t_\text{q}$ [Fig.\,\ref{Fig:quench}(c)]. Longer quenches are expected to destroy the prepared density profiles due to non-adiabaticity and the resulting tunneling dynamics when $a_\text{s}\approx0\,a_0$. As a measure of soliton stability we extracted the number of atoms in the central site after a hold duration of $50\,$ms. The data shows an exponential relationship between $N_\text{c}/N_\text{tot}$ and $t_\text{q}$ [Fig.\,\ref{Fig:quench}(c)]. For longer ramp durations ($t_\text{q}>20\,$ms) the wave packet spends significant time in the non-interacting regime, resulting in dispersion from a strong mismatch between the prepared and expected soliton density profiles. Fast quenches show increased stability with stable solitons forming when $t_\text{q}=1-5\,$ms. 


\section{Conclusion}

We have demonstrated versatile experimental techniques based on an optical accordion lattice for the controlled creation and study of bright matter-wave solitons in periodic potentials. A central component of this work is the precise calibration of the lattice parameters, measuring lattice depth and spacing over a range of $\dL=1.4(2)\,\upmu$m to 43.1(4)\,$\upmu$m and $V_0=0-100\,\Er$. In addition, we implemented microwave-assisted atom removal to achieve site-resolved preparation of initial states with well-defined occupation patterns. Furthermore, we demonstrated the use of repulsive solitons as an intermediate stage during the preparation process, eliminating the need to quench the trapping potential in addition to the scattering length. Our ability to prepare and manipulate initial states with high fidelity paves the way for the realization of more complex states in lattices \cite{trombettoni2001,franzosi2011}, such as gray and dark lattice solitons \cite{kevrekidis2003}, and for studying transport \cite{kivshar1993,rumpf2004} and interactions in tailored lattice geometries. 

Our results establish optical accordion lattices as versatile tools for preparing and investigating nonlinear phenomena in quantum gases. By generating the accordion lattice with multiple deflector frequencies, more complex lattice geometries, such as superlattices with several lattice constants, can be engineered, enabling studies of topological insulators \cite{cooper2019}. In particular, this approach could realize a Su–Schrieffer–Heeger chain, where topological solitons emerge from the system’s dimerized structure \cite{su1979,heeger1988}. Further extensions include generating higher-dimensional lattices \cite{baizakov2003} and synthetic gauge fields \cite{ye2018}, introducing controlled disorder \cite{sun2015,bai2024}, and creating bound states of lattice solitons \cite{kapitula2001}.

\vspace{3ex}

We acknowledge support by the EPSRC through a New Investigator Grant (EP/T027789/1), the Programme Grant ``Quantum Advantage in Quantitative Quantum Simulation'' (EP/Y01510X/1), and the Quantum Technology Hub in Quantum Computing and Simulation (EP/T001062/1). TH acknowledges funding from the European Research Council (ERC Starting Grant ``FOrbQ'', 101165353). 

The authors declare no conflicts of interest

The data used in this publication are openly available at the
University of Strathclyde Knowledge Base \cite{DataBase}.

\bibliography{./References.bib}

\end{document}